# Mixbiotic society measures:

# Assessment of community well-going as living system

Short title: mixbiotic society measures

Takeshi Kato[1*], Jyunichi Miyakoshi[1], Tadayuki Matsumura[2], Ryuji Mine[1], Hiroyuki Mizuno[2], Yasuo Deguchi[3]

[1] Hitachi Kyoto University Laboratory, Open Innovation Institute, Kyoto University, Kyoto, Japan

[2] Hitachi Kyoto University Laboratory, Center for Exploratory Research, Research & Development Group, Hitachi, Ltd., Tokyo, Japan

[3] Department of Philosophy, Graduate School of Letters, Kyoto University, Kyoto, Japan

* Corresponding author
E-mail: kato.takeshi.3u@kyoto-u.ac.jp (TK)

# Abstract

Social isolation is caused by the impoverishment of community (atomism) and fragmentation is caused by the enlargement of in-group (mobism), both of which can be viewed as social problems related to communication. To solve these problems, the philosophical world has proposed the concept of "mixbiotic society," in which individuals with freedom and diverse values mix and mingle to recognize their respective "fundamental incapability" each other and sublimate into solidarity. Based on this concept, this study proposes new mixbiotic society measures to evaluate dynamic communication patterns, replacing the conventional static social network analysis, with reference to classfication in cellular automata and particle reaction–diffusion that simulate living phenomena. Specifically, the hypothesis of measures corresponding to the four classes was formulated, and the hypothesis was validated by simulating the generation and disappearance of communication. As a result, considering communication patterns as multidimensional vectors, it found that the mean of Euclidean distance for "mobism," the variance of the relative change in distance for "atomism," the composite measure that multiplies the mean and variance of cosine similarity for "mixism," which corresponds to the well-going of mixbiotic society, and the almost zero measures for "nihilism," are suitable. Then, evaluating seven real-society datasets using these measures, we showed that the mixism measure is useful for assessing the livingness of communication, and that it is possible to typify communities based on plural measures. The measures established in this study are superior to conventional analysis in that they can evaluate dynamic patterns, they are simple to calculate, and their meanings are easy to interpret. As a future development, in conjunction with additional validation with various data sets, the mixbiotic society measures will be used to analyze issues of isolation and fragmentation, as well as in the fields of digital democracy and platform cooperativism, toward a desirable society.

## Introduction

Social isolation and fragmentation have become a global problem. Isolation and loneliness have a negative impact on health and well-being, and appropriate interventions and support for the target population are required [1, 2]. Fragmentation and polarization cause disturbances and conflicts in society, and effective facilitation to mediate intergroup relations is required [3, 4]. Social isolation and fragmentation are generally social issues related to communication, which are viewed as problems of community impoverishment (individual atomization: atomism) and hypertrophy (in-group crowding: mobism), respectively [5, 6].

To solve these social issues, the philosopher Deguchi proposes the concept of "mixbiotic society" that takes the symbiotic society a step further [7-9]. The "mixbiotic society" is a society in which individuals with freedom and diverse values mix and mingle in physical proximity to recognize their respective "fundamental incapability" each other and sublimate into solidarity. The "fundamental incapability" is that the individual "I" is incapable of any physical action alone, nor of complete control over others. The subject of mixbiotic society is not "Self as I," but "Self as WE" who are entrusted to each other. And "Self as WE" requires both openness (freedom) to avoid fragmentation and fellowship (solidarity) to avoid isolation. Deguchi also proposes "well-going" as a dynamic life as opposed to well-being as a static way of being.

Similar to mixbiotic societies, anthropologist Graeber and philosopher

Karatani also propose a vision of society in which the freedom of diverse individuals and the solidarity of community can be compatible without contradiction. Graeber holds up decentralization, equality, voluntary association, and mutual aid as core principles [10, 11]. As a mode of exchange, Karatani presents a system in which individual freedom and community mutual aid are compatible, while eliminating the negative aspects of the obligation to return in a gift economy [12, 13]. From the perspective of community or communication, in a mixbiotic society where freedom and solidarity are compatible, as shown by Deguchi, Graeber, and Karatani, the goal is to go beyond the dichotomy and aim for a medium between atomism and mobism, a well-going "mixism," so to speak.

According to sociologist Luhmann, a social system is an autopoietic system that forms emergent order through a network of processes of communication generation and disappearance [14, 15]. Autopoiesis is a theory of living systems proposed by biologists Maturana and Varela, a self-organizing system consisting of a recursive network of processes in which components produce components through interaction [16, 17]. From the perspective of autopoiesis, in order to achieve a living social system in which freedom and solidarity are in harmony, it is necessary to evaluate the generation and disappearance of communication and their networks, and to take measures to prevent atomism and mobism toward a well-going mixism.

Social network analysis is a well-known method for assessing the structure and meaning of social relationships in social systems [18, 19]. In social network analysis, a network is represented by a graph consisting of nodes (vertices) and links (edges) [20, 21], and the graph is analyzed using node-level and network-level statistical measures. Types of graphs include directed graphs, undirected graphs, and weighted graphs that represent the strength of link ties [22].

Node-level statistical measures mainly include degree, centrality, betweenness, closeness [23], eigenvector [24], and local clustering coefficient [25]. Network-level statistical measures mainly include diameter, average distance, average degree, density, global clustering coefficient [25], and directed graph reciprocity [20]. Cluster detection methods related to social fragmentation include hierarchical clustering by similarity [26], Girvan-Newman algorithm by betweenness [27], and modularity optimization [28]. However, these analysis methods are static and not sufficient from the viewpoint of living autopoiesis.

Dynamic evaluation methods include evolving network analysis and temporal network analysis [19]. Evolving network analysis examines degree distributions [29] and changes in average degree and cluster coefficient [30] to look at network growth and evolution. Temporal network analysis uses contact sequence graphs and interval graphs, where graphs are represented by nodes and links with time order, to look at disease transmission and information diffusion [31]. By introducing time-respecting paths [32] and strong connectivity [33], reachability, waiting time, and duration are analyzed. Also, by extending static measures, centrality, betweenness, closeness, adjacency correlation coefficient (time correlation coefficient) [34], motifs (subgraphs) [35], burstiness [36], etc. are analyzed.

According to network scientists Holme and Saramaki, the challenges of temporal network analysis include establishing a temporal network generation model that corresponds to the real world, elucidating the driving mechanism (why contact occurs when it does), and developing measures of temporal network structure that simply describe properties [31]. The first and second of these challenges require narrowing the scope of the social system and the communication objectives and fall outside the scope of this paper, which seeks to assess communication networks generally. The third is due to the fact that the existing measures are mostly extensions of static social network analysis measures. A more dynamic perspective is needed to analyze communication networks as living systems.

Therefore, we will take a break from network analysis and focus on cellular automata (CA), which are mathematical models that simulate living phenomena. In CA, a large number of cells laid out in a lattice in space change their own state in time according to an update rule as they interact with their neighbors [37, 38]. Computer scientist Wolfram showed that there are four classes of CA behavior [39]. Class 1 is a uniform state, class 2 is a periodic state, class 3 is a random and chaotic state, and class 4 is a mixture of order and randomness.

Computer scientist Langton named class 4 "the edge of chaos" and introduced a measure of complexity called the $\lambda$ parameter, showing that complexity increases from class 1 to 2, reaches a maximum in class 4, and decreases in class 3 [40]. Theoretical biologist Kauffman formed a hypothesis that life exists between order and chaos, or on "the edge of chaos" [41]. Theoretical biologist Gunji has shown that "the edge of chaos" can be expanded by adjusting the passivity and activity, synchrony and asynchrony of the cell update rules [42], and that the magnitude of the entropy of cell state and the variance of change of the entropy can be measures of "the edge of chaos," although not as clear as the $\lambda$ parameter [43, 44].

Viewed differently, CA behavior is the pattern of generation and disappearance of node states through interactions between nodes in a lattice network. Information scientist Miyakoshi has shown that four classes arise similar to CA in a mathematical model of particle reaction–diffusion (PRD), in which two types of particles are propagated between nodes of a two-dimensional lattice network based on reaction–diffusion equations [45]. The four classes are mapped onto a phase diagram with two axes of particle propagation speed and quantity, based on a decision formula consisting of the mean and range of the mutual information through computation time, and the AC component of the standard deviation of the particle quantity. The AC component corresponds to the variance of change of the entropy presented by Gunji. In class 4, the pattern formed by the two types of particles is life-like, between order and random.

With reference to the pattern formation of CA and PRD, to evaluate the dynamic patterns of generation and disappearance of communication in our target communication network, we should focus on the amount of the pattern change, the mutual amount before and after the change, and their variance or standard deviation. Considering that the amount of mutual information is based on information entropy, that the variance is square to the standard deviation, and that the AC component represents the magnitude of change, it is suggested that the variance of time-varying amount of communication patterns is useful as a measure of a mixbiotic society.

Based on the above, this study aims to develop a novel measure to dynamically evaluate communication networks toward a mixbiotic society where individual freedom and community solidarity are compatible. Then, based on this measure, we aim to obtain new guidelines for leading social systems toward living autopoiesis, that is to say, class 4 and the edge of chaos. The new measure will reflect a living autopoiesis perspective, or well-going, on the real society, as opposed to static social network analysis or evolving and temporal network analysis, which is primarily concerned with disease transmission and information diffusion. It is also better suited to the irregular communication networks of the real society than CA and PRD, which primarily target regular lattice networks.

Toward the above goal, this study first establishes a simple mathematical model to simulate the generation and disappearance of communication in a network that simulate a real community. Next, for the calculation results of this model, we will formulate several new measures of temporal changes in communication patterns, and examine what measures are suitable as mixbiotic society measures compared to the network analysis, in other words, what kind of measures can evaluate atomism, mobism, and mixism. Then, the new measures will be applied to seven datasets of real-society communication and their validity will be evaluated.

The remainder of this paper is organized as follows. The Methods section presents a mathematical model of communication and the calculation of hypotheses for mixbiotic society measures. In the first half of the Results section, simulations are performed based on the mathematical model, and the classes are mentioned together with the calculation results of the hypotheses, and the mixbiotic society measures are reset. In the second half of the section, we present the evaluation results with the new measures for several representative real-society datasets. In the Discussion section, we examine the validity and issues of the new measures and discuss its usefulness in comparison to social network analysis, CA and PRD. Finally, the Conclusions section presents overall conclusions and future perspectives.

# Methods

## Network model

In this section, we first formulate a simple mathematical model of the generation and disappearance of communication by forming a network that simulates a real community, then devise several hypothetical mixbiotic society measures that dynamically evaluate communication networks, and show how they are calculated.

Real-society networks are complex networks, but they are known to share common properties such as small-worldness and scale-freeness [46]. The Watts-Strogatz (WS) model [47] is a network model that simulates small-worldness, and the Barabási-Albert (BA) model [48] is prominent for scale-freeness. In the WS model, a regular graph is initially formed with equal degree $k$ for each of the $n$ vertices, and then the edges of the regular graph are randomly rewired with rewiring probability $p$. In the BA model, a complete graph is initially formed in which all $n_a$ vertices are joined to each other, then one vertex with $k$ edges is added at each step, and so on until the number of vertices reaches $n$. At each step, the $k$ edges are wired in proportion to the degree of the existing vertices.

The network graph $G$ with $n$ vertices formed by the WS and BA models is represented by the set of vertices $V$ shown in Eq (1) and the set of edges $E$ shown in Eq (2), or the adjacency matrix $A$ shown in Eq (3). Here, the graph $G$ is an undirected unweighted graph, the adjacency matrix $A$ is a symmetric matrix with zero diagonal components, and the components $a_{ij}\ (1 \leq i,j \leq n)$ are 0 or 1.

$$V = \{v_1, v_2, v_3, \cdots, v_n\}. \tag{1}$$

$$E = \{(v_1, v_2), (v_1, v_3), \cdots, (v_{n-1}, v_n)\}. \tag{2}$$

$$A = \begin{pmatrix} 0 & a_{12} & a_{13} & & a_{1n} \\ a_{21} & 0 & a_{23} & \cdots & a_{2n} \\ a_{31} & a_{32} & 0 & & a_{3n} \\ & \vdots & & \ddots & \vdots \\ a_{n1} & a_{n2} & a_{n3} & \cdots & 0 \end{pmatrix} \tag{3}$$

# Communication model

The role of the communication model in this paper is to provide computational results to validate the proposed hypothesis for mixbiotic society measures. As a mathematical model, the objective is to simulate the autopoiesis of a social system and to easily model the network of processes of generation and disappearance of communication.

The computation of the generation and disappearance of communication in a graph $G$ with $n$ vertices follows the flow shown in Fig 1. In this flow, roughly, the sender and receiver are chosen randomly from among the vertices according to the generation rate, information is sent from the sender to the receiver, and information is deleted randomly from among the vertices according to the disappearance rate. The states of isolation (atomism) and fragmentation (mobism) described in the Introduction section, as well as the state of mixism, the medium between the two, are expected to be reproduced depending on the setting of the generation and disappearance rates.

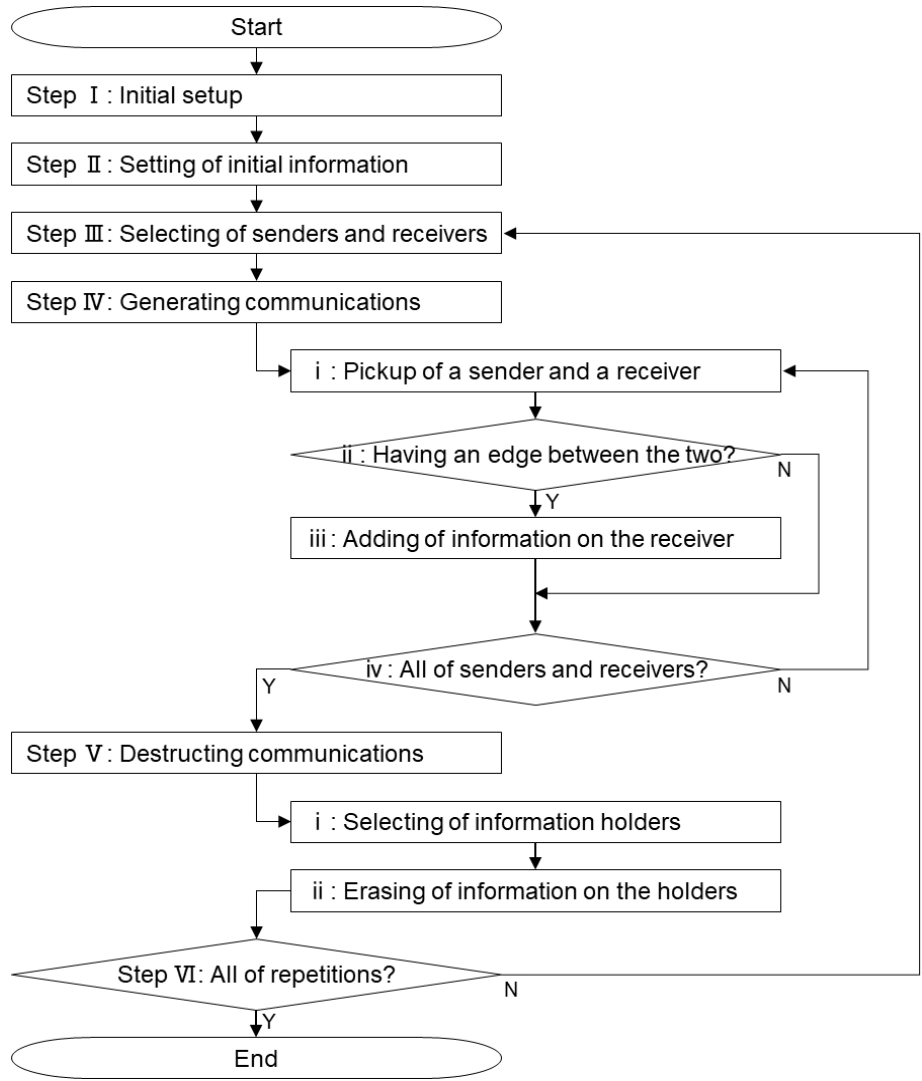

**Fig 1. Computational flow of communication model.**

First, as Step I (initial setup), we set the generation rate $g$ and disappearance rate $d$ of the communication, the information unit $u$ in the communication, the number of computational iterations $t_{max}$, and the set $Q$ of information that the vertices have, as shown in Eq (4). Here, we set the zero set $Q = 0$. Let $V_\emptyset$ be the set of vertices that have information (non-zero elements) corresponding to the information set $Q$. Here, $V_\emptyset = 0$.

$$Q = \{q_1, q_2, q_3, \cdots, q_n\},$$
$$V_\emptyset = \{v_i \in V \mid q_i \neq 0\}. \tag{4}$$

Next, as Step II (time $t = 0$), randomly select $n_0$ vertices from the vertex set $V$, giving information $u$ to $n_0$ vertices by operation $f_0(n_0, u)$ and leaving the elements of the other vertices 0, and rewrite the information set $Q$ and the vertex set $V_\emptyset$ as shown in Eq (5).

$$Q \xrightarrow{f_0(n_0, u)} Q,$$
$$V_\emptyset \rightarrow V_\emptyset = \{v_i \in V \mid q_i \neq 0\}. \tag{5}$$

In Step III, we count the number $n_\emptyset$ of vertices with non-zero information for the information set $Q$ at time $t$, calculate the number $n_s$ of vertices that can be the senders according to the generation rate $g$ as shown in Eq (6), and randomly select a set $V_s$ of possible senders from the vertex set $V_\emptyset$ by the operation $f_s(n_s)$. Similarly, as shown in Eq (7), calculate the number of vertices $n_r$ that can be the receivers according to the generation rate $g$, and randomly select a set $V_r$ of possible receivers from the vertex set $V$ by the operation $f_r(n_r)$. $Round$ is a rounding function.

$$n_s = Round[g \cdot n_\emptyset],$$
$$V_\emptyset \xrightarrow{f_s(n_s)} V_s. \tag{6}$$

$$n_r = Round[g \cdot n],$$
$$V \xrightarrow{f_r(n_r)} V_r. \tag{7}$$

In Step IV, one vertex $v_i$ and $v_j$ are taken from the sets $V_s$ and $V_r$, respectively, and the operation of adding information $u$ to the receiver vertex $v_j$ if the element $a_{ij}$ of the adjacency matrix $A$ is 1, as shown in Eq (8), rewriting the elements $q_j$ of information set $Q$ and vertex set $V_\emptyset$. Step IV is performed for all combinations of vertices $v_i$ and $v_j$.

$$\forall(v_i \in V_s, v_j \in V_r),$$

$$\text{if } a_{ij} = 1:$$

$$\text{then } Q \xrightarrow{q_j = q_j + u} Q,$$

$$\text{then } V_\emptyset \to V_\emptyset = \{v_i \in V \mid q_i \neq 0\}. \tag{8}$$

$$\text{else } Q = Q,$$

$$\text{else } V_\emptyset = V_\emptyset.$$

In Step V, we count the number $n_\emptyset$ of vertices with non-zero information again for the information set $Q$ rewritten by Step IV, calculate the number $n_d$ of vertices to erase information from the vertex set $V_\emptyset$ according to the disappearance rate $d$ as shown in Eq (9), randomly select a set $V_d$ of vertices to erase information from vertex set $V_\emptyset$ by operation $f_d(n_d, 0)$, and rewrite information sets $Q$ and $V_\emptyset$.

$$n_d = d \cdot n_\emptyset,$$

$$Q \xrightarrow{f_d(n_d, 0)} Q, \tag{9}$$

$$V_\emptyset \to V_\emptyset = \{v_i \in V \mid q_i \neq 0\}.$$

In Step VI, if time $t = t_{max}$, the calculation terminates; otherwise, it returns to Step III and repeats the calculation. At the end, the computation results in the information set $Q(t)$ and the vertex set $V_\emptyset(t)$ at each time $t$ from time 0 to $t_{max}$.

## Hypotheses for mixbiotic society measures

As noted in the Introduction section, the mixbiotic society measures require a perspective on the livingness of autopoiesis and the edges of chaos; with reference to the evaluation of classes in CA and PRD, it is promising to calculate some time-varying amount or its variance for the dynamic patterns of communication. On the other hand, the real society is not a regular network or regular interaction like CA and PRD, but a complex network or irregular interaction, and moreover, it does not always provide sufficient data like both. Therefore, rather than introducing complex decision formulas for a class of specific mathematical models like both, it is preferable to use a measure that is as easy to compute and interpret as possible, as stated by Holme and Saramaki.

Based on these considerations, as a hypothesis for the mixbiotic society measures, we focus on the change in the information set $Q$ between time $t$ and time $t+1$, and regard the communication pattern on the network consisting of $n$ vertices in the information set $Q$ as an $n$-dimensional vector. Then, we take the total information amount (sum of elements), Euclidean distance [49], and cosine similarity [50] of the $n$-dimensional vector as measures that are relatively easy to calculate and whose meaning is easy to interpret. These measures are familiar in the fields of document and image analysis. Although they are sometimes used in network analysis, Euclidean distance is mainly used for spatial representation of network history (e.g., [51, 52]) and cosine similarity for clustering of network structure and community detection (e.g., [53, 54]). The perspective of this paper, which focuses on temporal changes in communication patterns on the network with reference to CAs and PRDs rather than the spatial structure of the network, is novel.

Table 1 shows the hypotheses for the mixbiotic society measures. First, CA and PRD focus on large entropy and mutual information as class 3 features. Therefore, the mixbiotic society measure focuses on large total information amount change or distance as a feature of mobism corresponding to these two. Next, for large mutual information differences in small entropy as class 2 feature, the mixbiotic society measure focuses on large relative differences of distance as a feature of atomism.

Further, for large entropy variances and AC components as class 4 features (edge of chaos), the mixbiotic society measure focuses on large similarity variances as a feature of mixism. As for static class 1, the mixbiotic society measure regards it nihilism in which communication is nearly non-existent.

**Table 1. Hypotheses for mixbiotic society measures.**

|  | CA [42–44] | PRD [45] | Mixbiotic society measures | |
|---|---|---|---|---|
| Class 1 Static | Small entropy | All small value | All small value | Nihilism |
| Class 2 Sporadic | | Large difference of mutual information | Large relative difference of distance | Atomism |
| Class 4 Edge of chaos | Large entropy variance | Large AC component | Large similarity variance | Mixism |
| Class 3 Chaotic | Large entropy | Large mutual information | Large amount change or distance | Mobism |

The total information amount is the sum of the elements $q_i(t)$ of the information set $Q(t)$, and the change $I(t+1)$ between time $t$ and time $t+1$ is calculated from Eq (10). In the denominator, we took the case where all $n$ vertices have information unit $u$ as the reference. This measure looks at how the overall information has increased or decreased through communication.

$$I(t+1) = \frac{|\sum_{i=1}^{n} q_i(t+1) - \sum_{i=1}^{n} q_i(t)|}{n \cdot u}. \tag{10}$$

The Euclidean distance $L(t+1)$ is the distance between the vector of the information set $Q(t)$ at time $t$ and the vector of $Q(t+1)$ at time $t+1$, calculated from Eq (11). In the denominator, it is taken on the size of the $n$-dimensional unit vector. The relative change in Euclidean distance, $L_R(t+1)$, is calculated from Eq (12). The denominator is the magnitude of the $Q(t+1)$ vector. These measures look at how far apart the communication states are.

$$L(t+1) = \frac{\sqrt{\sum_{i=1}^{n}(q_i(t+1) - q_i(t))^2}}{\sqrt{n} \cdot u}. \tag{11}$$

$$L_R(t+1) = \frac{\sqrt{\sum_{i=1}^{n}(q_i(t+1) - q_i(t))^2}}{\sqrt{\sum_{i=1}^{n} q_i(t+1)^2}}. \tag{12}$$

The cosine similarity $S(t+1)$ is the similarity between the vector of $Q(t)$ at time $t$ and the vector of $Q(t+1)$ at time $t+1$, calculated from Eq (13). Cosine similarity generally takes a value between $-1$ and $1$, but here it is normalized to a value between $0$ and $1$ since $q_i(t+1) \cdot q_i(t) \geq 0$. This measure looks at how similar the communication states are. The cosine similarity $S(t+1)$ is a measure about the direction (angle) of two vectors and is complementary to the Euclidean distance $L(t+1)$. Cosine similarity is also closely related to the correlation coefficient.

$$S(t+1) = \frac{\sum_{i=1}^{n} q_i(t+1) \cdot q_i(t)}{\sqrt{\sum_{i=1}^{n} q_i(t+1)^2} \cdot \sqrt{\sum_{i=1}^{n} q_i(t)^2}}. \tag{13}$$

The mean $\mu_I$ and variance $\sigma_I^2$ of the total information change $I(t)$, the mean $\mu_L$ and variance $\sigma_L^2$ of the Euclidean distance $L(t)$, the mean $\mu_{LR}$ and variance $\sigma_{LR}^2$ of its relative change $L_R(t)$, and the mean $\mu_S$ and variance $\sigma_S^2$ of the cosine similarity $S(t)$ are calculated from Eqs (14), (15), (16), and (17) each other, as well as general calculation formulas. Unbiased variance was used for the variance. The reason for obtaining these means and variances is that they follow the use of means and variances of changes (or AC components) in CA and PRD class decision.

$$\mu_I = \sum_{t=1}^{t_{max}} I(t),$$

$$\sigma_I^2 = \frac{1}{t_{max}-1} \sum_{t=1}^{t_{max}} (I(t) - \mu_I)^2.$$

(14)

$$\mu_L = \sum_{t=1}^{t_{max}} L(t),$$

$$\sigma_L^2 = \frac{1}{t_{max}-1} \sum_{t=1}^{t_{max}} (L(t) - \mu_L)^2.$$

(15)

$$\mu_{LR} = \sum_{t=1}^{t_{max}} L_R(t),$$

$$\sigma_{LR}^2 = \frac{1}{t_{max}-1} \sum_{t=1}^{t_{max}} (L_R(t) - \mu_{LR})^2.$$

(16)

$$\mu_S = \sum_{t=1}^{t_{max}} S(t),$$

$$\sigma_S^2 = \frac{1}{t_{max}-1} \sum_{t=1}^{t_{max}} (S(t) - \mu_S)^2.$$

(17)

## Results

### Network formation

First, the WS [47] model and the BA model [48] are used to form a network that simulates a community. The number of vertices $n$ of both networks, i.e., the number of people constituting the community, was set to 100, referring to the Dunbar number (the upper limit of the number of people for which a group can maintain a stable

social state, 100–230 [55, 56]).

For the WS model with small-worldness, the network was formed by starting with a regular graph with $n = 100$ vertices and $k = 4$ degrees per vertex, and rewiring edges with probability $p = 0.7$. The rewiring probability $p = 0.7$ is due to the fact that all of the findings for various actual networks were generally 0.7 [57]. $k = 4$ is a relatively small value in the survey results. This is because we thought that a larger $k$ would make it more difficult to reproduce the state of isolation (atomism). For the BA model with scale-freeness, we started with a complete graph with $n_a = 3$ vertices and added $k = 2$ edges until we reached $n = 100$ vertices, forming the network. The reason for starting with $n_a = 3$ was to roughly align the WS model with the total degree $n \cdot k/2 = 200$.

Fig 2 shows the results of network graph formation for the WS and BS models, and Table 2 shows the graph features. With respect to the degree distribution in Fig 2, the WS model has a symmetric distribution, while the BA model has a power-law distribution. With respect to graph features, the WS model has a longer diameter and mean distance and a smaller mean cluster coefficient than the BA model.

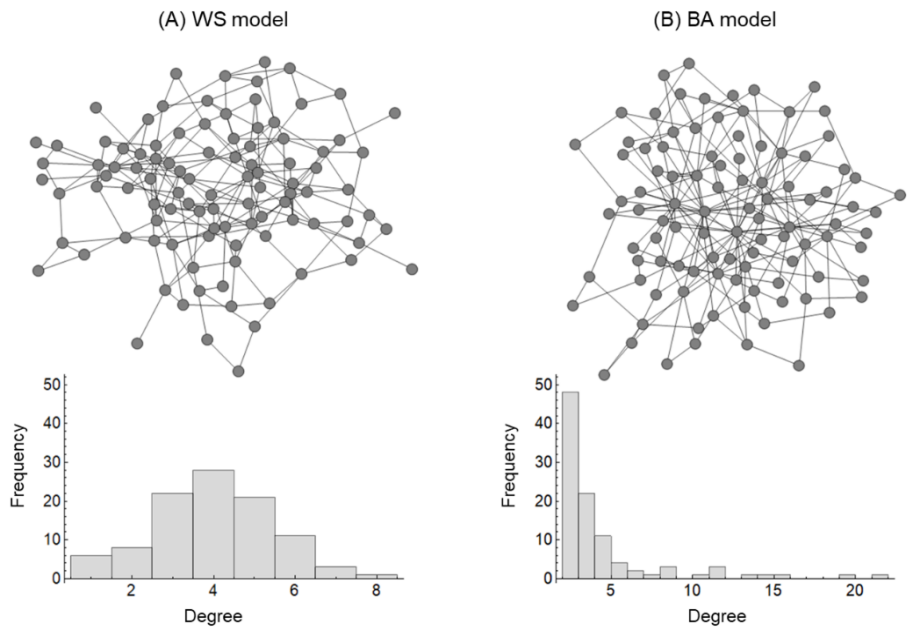

**Fig 2. Network formation results.**

(A) WS model, and (B) BA model.

**Table 2. Network graph features of WS model and BA model.**

|  | WS model | BA model |
|---|---|---|
| Vertex count | 100 | 100 |
| Edge count | 200 | 197 |
| Diameter | 7 | 5 |
| Mean distance | 3.62 | 3.00 |
| Density | 0.0404 | 0.0398 |
| Mean Clustering Coefficient | 0.059 | 0.118 |

# Communication simulation

In the communication model, the communication generation rate $g$ and disappearance rate $d$ were used to calculate as parameters for the networks of the WS and BA models, respectively. Specifically, a mesh of total $140$ points was constructed by adding $19$ points in the neighborhood of $g \sim d$ to the $11 \times 11 = 121$ points that were changed between $g, d = 0 \sim 1$ in $0.1$ increments. For each point of the mesh, the computational flow shown in Fig 1 was flowed between time $t = 0 \sim t_{max}$ in $1$ increments to obtain the time average of the hypotheses for mixbiotic society measures in the Methods section. In addition, $100$ trials of this flow were repeated to obtain the average of the number of times. As initial settings for Steps I and II, we set the information unit $u = 1$, the number of calculation iterations $t_{max} = 100$, and the number of vertices $n_0 = 10$, in common for the WS and BA models.

Fig 3 shows examples of communication patterns for the WS and BA models. In each of Figs 3A–3D, the value of element $q_i(t)$ of the information set $Q(t)$ in the computational flow of Fig 1 is represented by the red color density of the vertices on the network. Figs 3A and 3B are examples of patterns at times $t = 30$ and $31$ for the WS model with $g = 0.4, d = 0.3$, and Figs 3C and 3D are examples of patterns at times $t = 70$ and $71$ for the BA model with $g = 0.8, d = 0.6$. For each calculation flow, the value of the measures between time $t$ and time $t + 1$ is calculated for such a pattern using Eqs (10)–(13), the time average and variance from time $t = 0$ to $t_{max}$ are calculated using Eqs (14)–(17), and the calculation flow is repeated to obtain the trial count average.

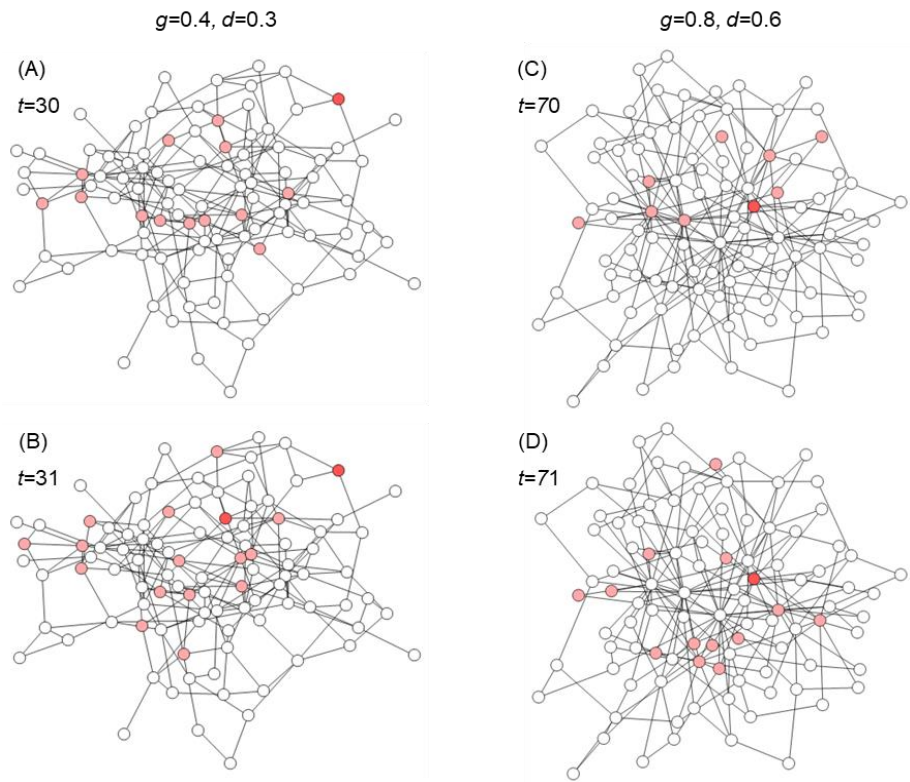

**Fig 3. Examples of communication patterns.**

(A) WS model: $g = 0.4, d = 0.3, t = 30$. (B) $t = 31$. (C) BA model: $g = 0.8, d = 0.6, t = 70$. (D) $t = 71$.

Figs 4 and 5 show the calculation results of the hypotheses for mixbiotic society measures of the WS and BA models, respectively. Figs 4A and 5A show the mean $\mu_I$ of the change in total information amount, Figs 4B and 5B show the variance $\sigma_I^2$, Figs 4C and 5C show the mean $\mu_L$ of Euclidean distance, Figs 4D and 5D show the variance $\sigma_L^2$, Figs 4E and 5E show the mean $\mu_{LR}$ of relative change in Euclidean distance, Figs 4F and 5F show the variance $\sigma_{LR}^2$, Figs 4G and 5G show the mean $\mu_S$ of cosine similarity, and Figs 4H and 5H show the variance $\sigma_S^2$. In Figs 4A–4H and 5A–5H, respectively, the x-axis is the communication generation rate $g$, the y-axis is the disappearance rate $d$, and the z-axis is the value of mixbiotic society measure.

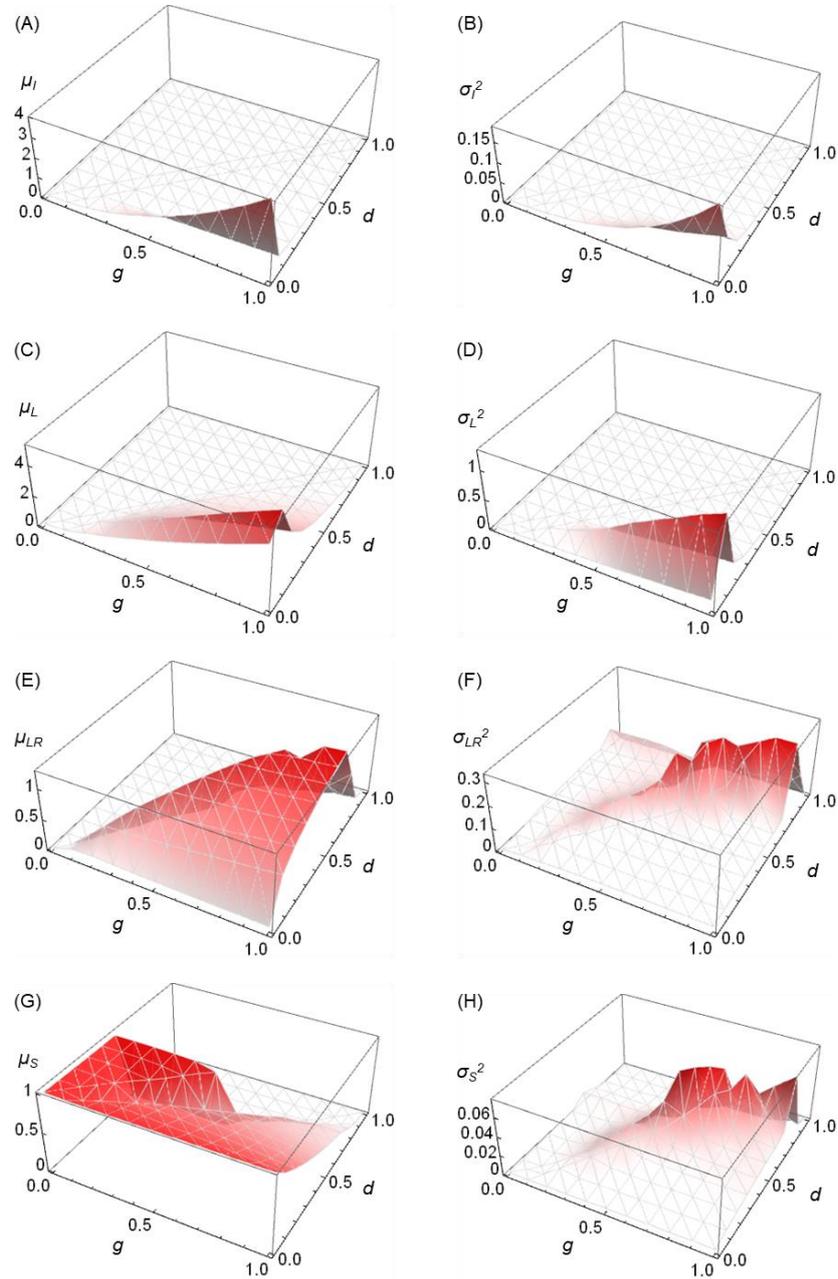

**Fig 4. Calculation results on WS model for hypotheses of mixbiotic society measures.**

(A) Mean $\mu_I$ of change in total information amount. (B) Variance $\sigma_I^2$. (C) Mean $\mu_L$ of Euclidean distance. (D) Variance $\sigma_L^2$. (E) Mean $\mu_{LR}$ of relative change in

Euclidean distance. (F) Variance $\sigma_{LR}^2$. (G) Mean $\mu_S$ of cosine similarity. (H) Variance $\sigma_S^2$.

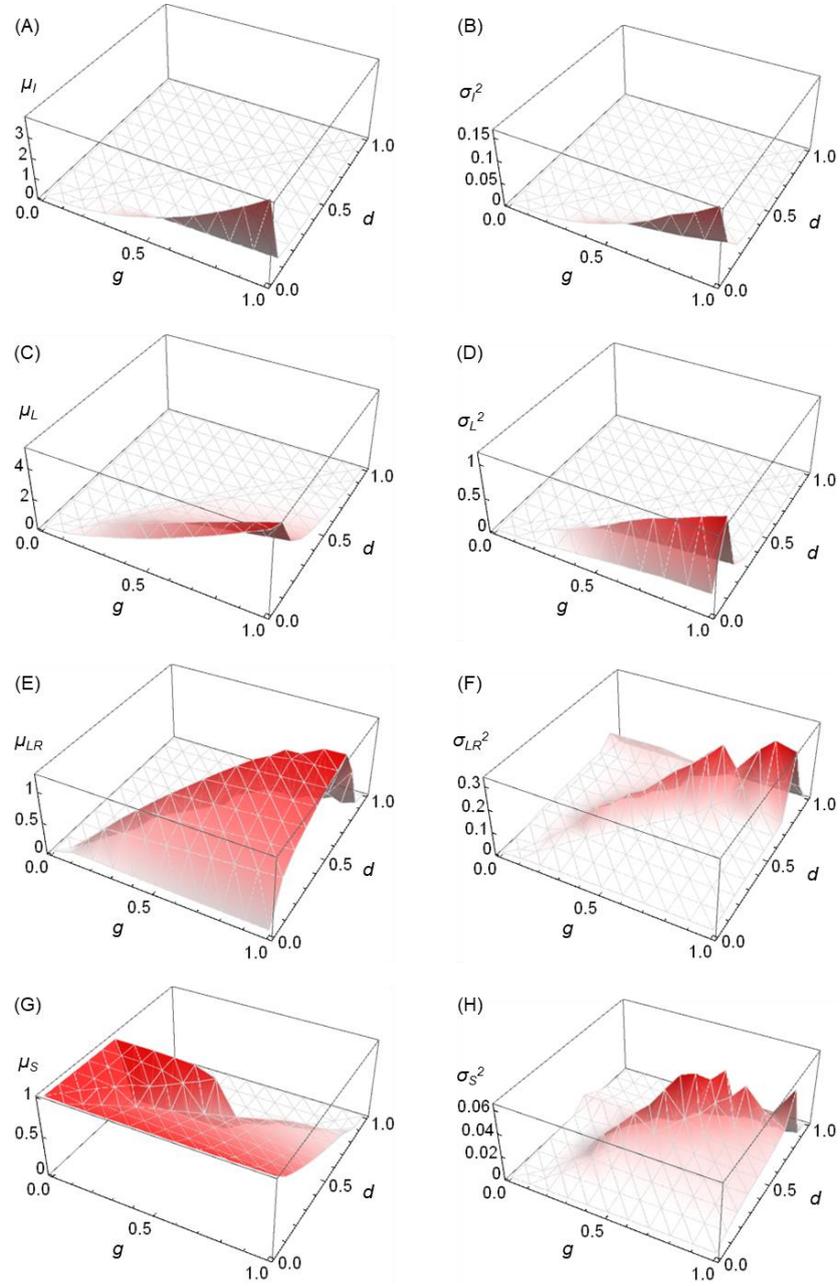

**Fig 5. Calculation results on BA model for hypotheses of mixbiotic society measures.** (A) Mean $\mu_I$ of change in total information amount. (B) Variance $\sigma_I^2$. (C)

Mean $\mu_L$ of Euclidean distance. (D) Variance $\sigma_L^2$. (E) Mean $\mu_{LR}$ of relative change in Euclidean distance. (F) Variance $\sigma_{LR}^2$. (G) Mean $\mu_S$ of cosine similarity. (H) Variance $\sigma_S^2$.

First, let us look at the calculation results of the WS model in Fig 4. Figs 4A and 4B show that the mean $\mu_I$ and variance $\sigma_I^2$ of the change in the total information amount become larger as the generation rate $g$ increases and the disappearance rate $d$ decreases. This trend is particularly pronounced for $g > 0.6$ and $d < 0.3$.

Figs 4C and 4D show that the mean $\mu_L$ and variance $\sigma_L^2$ of Euclidean distance become larger as $g$ becomes larger and $d$ becomes smaller, as in Figs 4A and 4B. The trend of $\mu_L$ is not as extreme as that of $\mu_I$ and $\sigma_I^2$. As a measure $M_{mob}$ representing the tendency of mobism (enlargement and crowding of the in-group) for $g$ and $d$, $\mu_L$ is more suitable than $\mu_I$ and $\sigma_I^2$. The peak of $\sigma_L^2$ occurs near $d \sim 0.1$. The reason for this is that at $d = 0$, all senders always transmit information and all receivers do not disappear, so the variation in the state of communication is scarce.

Figs 4E and 4F show that the mean $\mu_{LR}$ and variance $\sigma_{LR}^2$ of the relative change in Euclidean distance become larger as approaching the diagonal that is $g \sim d$, but the change in $\mu_{LR}$ is more gradual than the change in $\sigma_{LR}^2$. The region of $d > 0.7$, where $\sigma_{LR}^2$ is particularly large, is the region where the cosine similarity $\mu_S$ in Fig 4G is small, indicating that communication, if it occurs, is sporadic. Therefore, $\sigma_{LR}^2$ is more suitable than $\mu_{LR}$ as a measure of atomism (community impoverishment and individual isolation), $M_{atom}$.

In Figs 4G and 4H, the mean $\mu_S$ of the cosine similarity is large in the region $d < 0.5$. The reason why $\mu_S$ is particularly large in the regions of $g < 0.5$ and $d <$

0.5 is that the initial state is relatively well maintained due to less communication generation and disappearance. The variance $\sigma_S^2$ is similar in trend to the variance $\sigma_{LR}^2$ in Fig 4F. The hypotheses of the mixbiotic society measures in Table 2 expected that the variance $\sigma_S^2$ would represent the feature of mixism (edge of chaos) between atomism (sporadic) and mobism (chaotic), but judging from Fig 4H, it is not sufficient.

Note that in Figs 4F and 4H, the three-dimensional graphs are serrated. This tendency did not change even after the calculations were redone. The reason for the serration is presumably due to the irregularity of the network structure and the discreteness of the number of vertices, the information units, and the calculation method, in addition to the effect of the mesh partitioning. Since elaboration of the calculation is not the main purpose of this paper, we will not go into it further.

Next, the calculation results of the BA model in Fig 5 are compared with those of the WS model in Fig 4; although there are differences between the BA and WS models in terms of network structure, degree distribution, diameter, mean distance, and cluster coefficients, the calculation results of both models are almost the same. This indicates that $\mu_L$ as the measure $M_{mob}$ for mobism and $\sigma_{LR}^2$ as the measure $M_{atom}$ for atomism are suitable, as explained in Fig 4. Comparing Figs 5F and 5H, the variance of the cosine similarity $\sigma_S^2$ does not represent a difference from atomism and is still insufficient as a measure of mixism, and the hypotheses presented in Table 2 should be revised.

Therefore, based on the results of Figs 4H and 5H, we reconsider the mixbiotic society measure representing mixism. In mixism, the communication patterns are considered to fluctuate between atomism (sporadic) and mobism (chaotic) in living way, balancing similarity and dissimilarity. Therefore, we shall reset the composite measure

$\mu_S \cdot \sigma_S{}^2$, which is the variance $\sigma_S{}^2$ indicating dissimilarity focused on in Table 2, multiplied by the mean $\mu_S$ indicating similarity.

Fig 6 shows the calculation results of the composite measure $\mu_S \cdot \sigma_S{}^2$ in the WS and BA models. $\mu_S \cdot \sigma_S{}^2$ is large in the region near $g \sim 0.5$ and $d \sim 0.5$ on the diagonal line, which is $g \sim d$. In this region, the generation and disappearance of communication are moderately balanced between the two. This indicates that communication is neither isolating nor crowding, but rather moderately persistent, i.e., well-going. Therefore, instead of the hypothesis $\sigma_S{}^2$ shown in Table 2, $\mu_S \cdot \sigma_S{}^2$ is more suitable as a measure $M_{mix}$ for mixism.

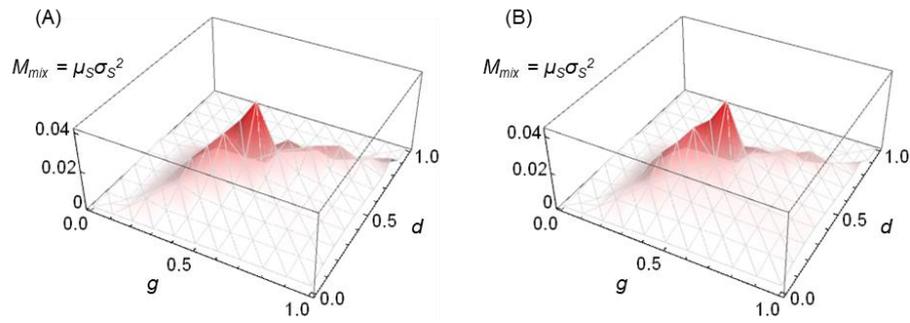

**Fig 6. Calculation results for composite measure $\mu_S \cdot \sigma_S{}^2$ of mixism.**
(A) WS model, and (B) BA model.

Based on the results from Figs 4 to 6, Fig 7 shows the phase diagram synthesizing the results of atomism $M_{atom} = \sigma_{LR}{}^2$, mixism $M_{mix} = \mu_S \cdot \sigma_S{}^2$, and mobism $M_{mob} = \mu_L$ in the WS and BA models. Each value was normalized by its maximum value and represented as a phase diagram by superimposing the three 3D graphs and viewing them from the infinity of the z-axis; in the region $d > g$ for the diagonal of $g \sim d$, communication does not persist but almost disappears ($M_{atom,mix,mob} \approx 0$), and this is shown as nihilism.

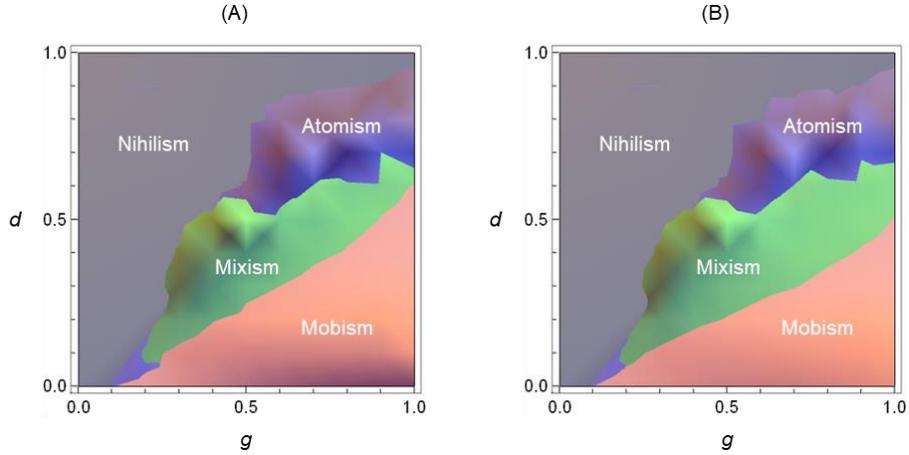

**Fig 7. Phase diagram of mixbiotic society measures.**

(A) WS model, and (B) BA model.

Fig 7 shows that the regions indicated by the measures of atomism $M_{atom} = \sigma_{LR}^2$, mixism $M_{mix} = \mu_S \cdot \sigma_S^2$, and mobism $M_{mob} = \mu_L$ and the region for nihilism $M_{atom,mix,mob} \approx 0$ are well segregated from each other. Therefore, it would be useful to use these as mixbiotic society measures. The region of mixism is at the edge of mobism and does not fall into atomism or nihilism. This leads to livingness dwelling on the edge of chaos. Note that the region in Fig 7B is wider than the region of mixism in Fig 7A, but the widened area is the region where the values of $M_{atom}$, $M_{mix}$, and $M_{mob}$ are small, and the difference is not remarkable.

As supporting results, Fig 8 shows typical examples of communication trajectories in the four phases of nihilism, atomism, mixism, and mobism. Figs 8A–8D are examples of the WS model and Figs 8E–8H are examples of the BA model. In the polar coordinate system, the magnitude of the $n$-dimensional vector of the information set $Q(t)$ is taken as the dynamic radius $r$, as shown in Eq (18), and the angle formed by the $n$-dimensional vector and the unit vector $\mathbf{1}$ as the declination angle $\theta$, as shown in Eq (19). The trajectory was drawn by connecting the points at each time from $t = 0$ to $t_{max}$ in turn.

$$r = \sqrt{\sum_{i=1}^{n} q_i(t)^2}. \tag{18}$$

$$\theta = \text{Arccos}\left(\frac{\sum_{i=1}^{n} q_i(t) \cdot 1}{\sqrt{\sum_{i=1}^{n} q_i(t)^2} \cdot \sqrt{\sum_{i=1}^{n} 1^2}}\right). \tag{19}$$

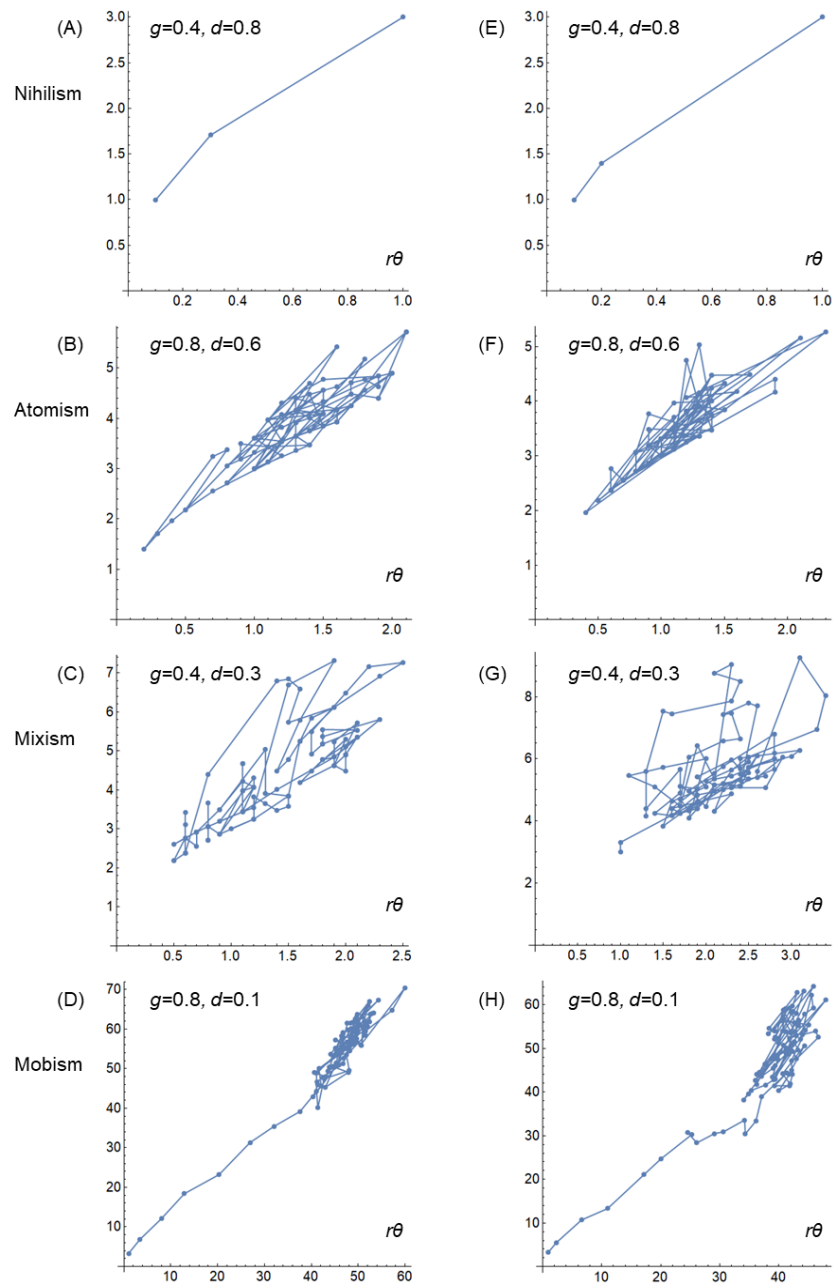

**Fig 8. Communication trajectories in four phases: nihilism, atomism, mixism, and mobism.**

(A)–(D) WS model, and (E)–(H) BA model.

In the examples of nihilism in Figs 8A and 8E ($g = 0.4, d = 0.8$), communication does not persist and quickly disappears. This is a quiet nihility, a state of death without dynamic livingness.

In the atomism examples in Figs 8B and 8F ($g = 0.8, d = 0.6$), at first glance it appears that the trajectories are connected and persistent, but when considered with the large similarity variance in Figs 4F and 5F, the communication is isolated, occurring only in a sporadic fashion. In addition, because many vertices in the atomism repeatedly become a sender of transmission and many information disappear, the orbit oscillates significantly in certain directions corresponding to the large value of $\sigma_{LR}^2$. This shows a rambling and inharmonic situation.

The examples of mixism in Figs 8C and 8G ($g = 0.4, d = 0.3$) are characterized by a larger orbit declination $\theta$ than that of atomism and mobism, because they combine dissimilarity as well as similarity, corresponding to a larger $\mu_S \cdot \sigma_S^2$. And the orbits do not stay in a particular direction, but go around circularly. This situation indicates that communication is in a state of well-going and is mixed. The trajectory of mixism is not as clear as the trajectory of attractor at the edge of chaos, but it seems to suggest that the mixism corresponds to class 4. In the present communication model, generation and disappearance were random, but it is expected that the introduction of some rules would bring the trajectory closer to an attractor-like trajectory.

In the mobism examples in Figs 8D and 8H ($g = 0.8, d = 0.1$), many vertices become the sender of transmission and continue to send information constantly, and information is stored at each vertex, so the orbit extends roughly in the direction of the unit vector declination angle $\theta$, corresponding to a large $\mu_L$. After a certain degree of elongation, the trajectory stays in a certain region because disappearance occurs (by the way, if $d = 0$, the trajectory continues to elongate). Note that there are two possible meanings for the situation in which the trajectory extends in a particular direction. Although we have not gone into the content of communication in this communication model, if the content is similar (not the similarity of patterns shown in Figs 4G and 5G), it means enlargement and crowding of the in-group, and if the content is dispersed and different, it means chaos and confusion. Neither situation is desirable for communication.

Note that the illustration of the trajectories of communication patterns on the network using polar coordinates is considered useful as a visual capture of the results of the mixbiotic society measures. Based on the above results from Figs 4 to 8, the hypotheses presented in Table 2 are reviewed, and the reset mixbiotic society measures are summarized in Table 3.

**Table 3. Mixbiotic society measures reset based on communication simulations.**

| Phase/class | Meaning | Formula | Feature |
|---|---|---|---|
| Nihilism | All small value | $M_{atom,mix,mob} \approx 0$ | Static and silent |
| Atomism | Large relative difference of distance | $M_{atom} = \sigma_{LR}^2$ | Sporadic and isolate |
| Mixism | Balance of similarity and dissimilarity | $M_{mix} = \mu_S \cdot \sigma_S^2$ | Living and cyclic |
| Mobism | Large distance | $M_{mob} = \mu_L$ | Chaotic or uniform |

## Validation with real-society data sets

In order to validate the mixbiotic society measures, we performed the calculations here on seven datasets of temporal networks with potentially different communication pattern features. As datasets, we used the network data repository collected by Rossi and Ahmed [58] and the face-to-face contact datasets jointly collected by the ISI Foundation, the CNRS (Centre national de la recherche scientifique), and Bitmanufactory [59].

Of the seven datasets used here, the first is contact data between students in a French high school measured using RFID sensors [60, 61]. Contact data are considered to approximate face-to-face communication [62]; the second is contact data between children and teachers in a French primary school [63, 64]; the third is contact data between individuals at workplace in a French office building [65, 66]; the fourth is contact data between individuals including adults, adolescents, and children in a village rural Malawi [67, 68]; the fifth is contact data between participants measured using radio badges at the Hypertext 2009 conference [69, 70]; the sixth is data on messages sent and received in the online community for students at the University of California, Irvine [71, 72]; the seventh is data on email sent and received by the U.S. Democratic National Committee [73, 74].

Table 4 shows the graph features of the network over all time in the seven data sets. $t_{count}$ is the number of data at the time recorded in each dataset, and $t_{max}$ is the number of data counting the same time as 1, in the same usage as $t_{max}$ in Eqs (10)–(17). ∞ in the table indicates that there were vertices that were not connected throughout all time. Table 4 shows that the online community and Email have smaller graph density and mean cluster coefficients than the other five, and that the ratio of $t_{count}$ to $t_{max}$ is close to 1, indicating that there is not much co-temporal communication. The figures of the network graphs except for the village can be found in the literature [60][63][65][69][71][73], respectively.

**Table 2. Network graph features of seven datasets.**

| | High school [60, 61] | Primary school [63, 64] | Workplace [65, 66] | Village [67, 68] | Conference [69, 70] | Online community [71, 72] | Email [73, 74] |
|---|---|---|---|---|---|---|---|
| Vertex count | 327 | 242 | 217 | 86 | 113 | 1,899 | 1,891 |
| Edge count | 5,818 | 8,317 | 4,274 | 347 | 2,498 | 22,195 | 5,598 |
| Diameter | 4 | 3 | 5 | ∞ | 3 | ∞ | ∞ |
| Mean distance | 2.159 | 1.732 | 1.882 | ∞ | 1.656 | ∞ | ∞ |
| Density | 0.1092 | 0.2852 | 0.1824 | 0.0949 | 0.3470 | 0.0077 | 0.0025 |
| Mean Clustering Coefficient | 0.504 | 0.526 | 0.381 | 0.527 | 0.535 | 0.109 | 0.209 |
| $t_{count}$ | 188,508 | 125,773 | 78,439 | 102,293 | 20,818 | 61,734 | 39,264 |
| $t_{max}$ | 7,375 | 3,100 | 18,488 | 43,438 | 5,246 | 60,774 | 21,751 |

Table 5 and Fig 9 show the results of the measure calculations according to the method described in Eqs (10)–(17). In Table 5, the values of mixism for the WS and BA models (for $g = 0.4, d = 0.3$, the same as in Figs 8C and 8G) are included for comparison. In Fig 9, for ease of understanding, we drew a radar chart with six typical cases: the WS model and high school (dark and light green), workplace and village (dark and light orange), and online community and email (dark and light blue). In this chart, for ease of reading, the values for each item are normalized using the maximum value.

**Table 5. Calculation results on seven datasets for mixbiotic society measures.**

| | WS model mixism | BA model mixism | High school | Primary school | Workplace | Village | Conference | Online community | Email |
|---|---|---|---|---|---|---|---|---|---|
| $\mu_I$ | 0.0152 | 0.0253 | 0.0213 | 0.0546 | 0.0118 | 0.0176 | 0.0234 | 0.0000 | 0.0014 |
| $\sigma_I^2$ | 0.0002 | 0.0004 | 0.0004 | 0.0021 | 0.0002 | 0.0004 | 0.0009 | 0.0000 | 0.0000 |
| $\mu_L$ ($M_{mob}$) | 0.3300 | 0.4253 | 0.3266 | 0.6673 | 0.1560 | 0.1506 | 0.2171 | 0.0411 | 0.0750 |
| $\sigma_L^2$ | 0.0113 | 0.0164 | 0.0133 | 0.0254 | 0.0091 | 0.0130 | 0.0235 | 0.0002 | 0.0109 |
| $\mu_{LR}$ | 0.7682 | 0.7916 | 0.7394 | 0.9128 | 0.8151 | 0.6463 | 0.7954 | 1.2431 | 1.6112 |
| $\sigma_{LR}^2$ ($M_{atom}$) | 0.0672 | 0.0515 | 0.0253 | 0.0236 | 0.2070 | 0.2652 | 0.2648 | 0.1716 | 4.1790 |
| $\mu_S$ | 0.7032 | 0.6865 | 0.7287 | 0.5933 | 0.6439 | 0.7158 | 0.6431 | 0.1594 | 0.1573 |
| $\sigma_S^2$ | 0.0210 | 0.0187 | 0.0108 | 0.0114 | 0.0722 | 0.0864 | 0.0860 | 0.0931 | 0.0636 |
| $\mu_S \cdot \sigma_S^2$ ($M_{mix}$) | 0.0149 | 0.0130 | 0.0078 | 0.0067 | 0.0465 | 0.0619 | 0.0553 | 0.0148 | 0.0100 |

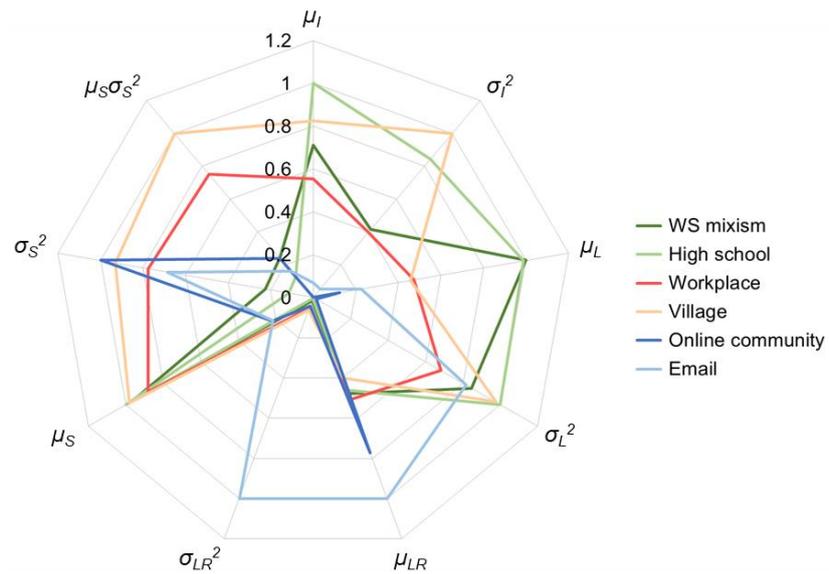

**Fig 9. Rader chart of calculation results for mixbiotic society measures.**

In Fig 9, the mixism of the WS model and the high school are more similar than the other four. This is because this communication model was modeled such as communication is ongoing, and in high school, relatively similar members always gather in the same place to communicate, which gives rise to similar communication patterns. As well as these two, the mixism of BA model and the primary school show the same tendency. However, in high schools and primary schools, the variance of similarity $\sigma_S^2$ and the measure of mixism $\mu_S \cdot \sigma_S^2$ are considered to be small because the members are relatively fixed. Note that although $\mu_L$ appears relatively large, this is not a situation of mobism as shown in Figs 8D and 8H.

In the workplace and village, both the similarity $\mu_S$ and its variance $\sigma_S^2$ are large, and therefore the measure of mixism $\mu_S \cdot \sigma_S^2$ is larger than those of the other four models. Although the mixism of the WS and BA models had larger $\mu_S \cdot \sigma_S^2$ in the simulations conducted in this study, the workplace and village are relatively larger than them, and are even more mixism-like and mixbiotic society-like. The reason for this is inferred to be that both $\mu_S$ and $\sigma_S^2$ were larger in the workplace and village because the same members gathered according to the time of day and then gathered with other members when the time of day changed, whereas the generation and disappearance of communication was random in the present model (see [68] for the tendency of temporal activity in village). Note that conferences also have the same tendency as workplace and village, as participants gather and depart. The fact that $\mu_S \cdot \sigma_S^2$, which balances the generation and disappearance of communication, is large in workplace, village, and conference, where communication is valued, suggests that $M_{mix} = \mu_S \cdot \sigma_S^2$ is useful as a composite measure of mixism.

Online community and email have a peculiar shape compared to the other four. The small similarity $\mu_S$ and the large relative change in distance $\mu_{LR}$ and variance of similarity $\sigma_S^2$ indicate that the instantaneous change in communication patterns is large, i.e., the communication is sporadic, as explained in Table 4, and is atomism-like. The reason why the variance $\sigma_{LR}^2$ is particularly larger for email may be that the communication has even less context over time than in online community, and the tendency of atomism is more stronger. Note that although $\sigma_{LR}^2$ was set as a measure of atomism in Table 3, $\mu_{LR}$ may be more appropriate in Fig 9. Additional validation of atomism is a subject for future study.

Fig 10 shows the communication trajectories drawn using the same method as in Fig 8. Figs 10A–10F show high school, primary school, workplace, village, online community, and email, respectively.

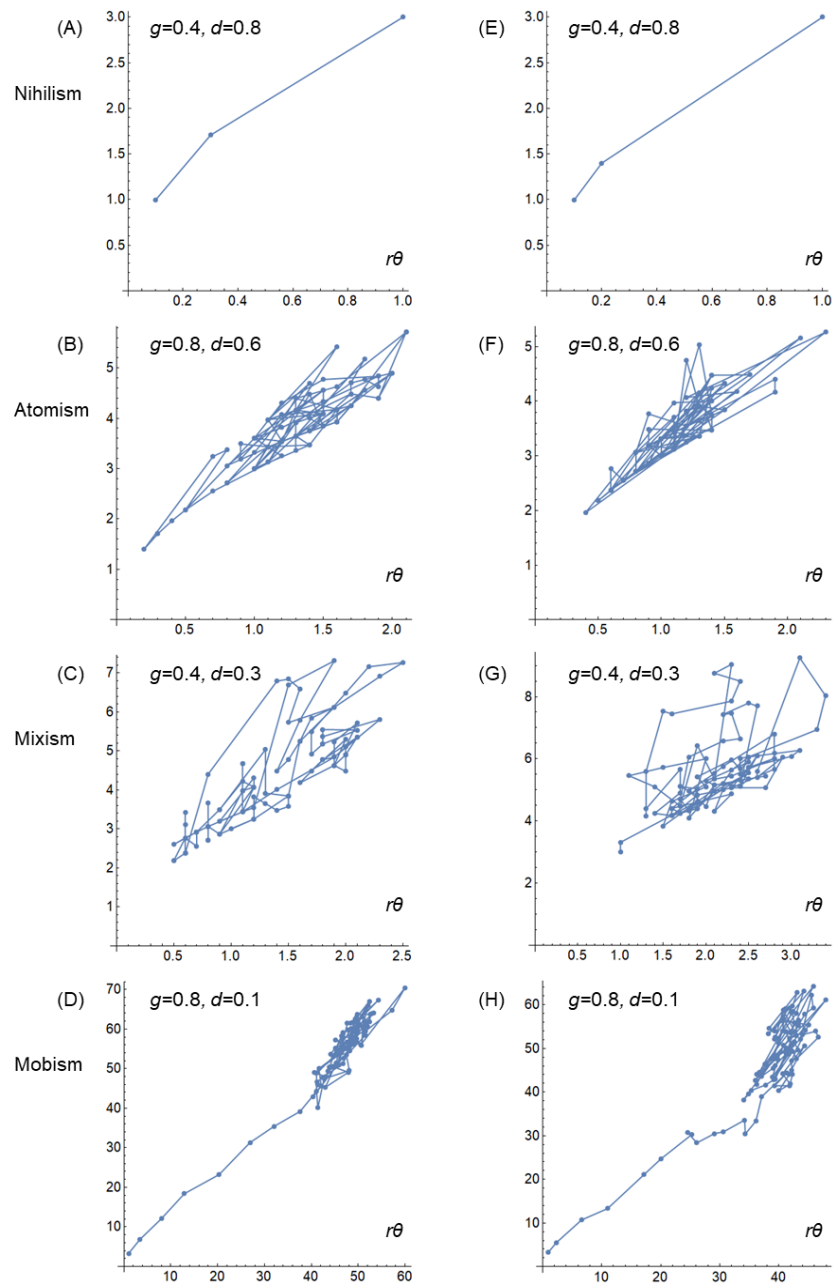

**Fig 10. Communication trajectories.**

(A) high school, (B) primary school, (C) workplace, (D) village, (E) online community, and (F) email.

Although it is difficult to see because the trajectories were drawn for all times and overlap, the workplace in Fig 10C and the village in Fig 10D show aspects of a mixism, with cyclical trajectories similar to those shown in Figs 8C and 8G in some places. The high school in Fig 10A and the primary school in Fig 10B are special environments where, unlike mixism, collective action takes place.

The online community in Fig 10E and email in Fig 10F show occasional trajectories that extend to the upper right of the figure due to events such as simultaneous transmissions, but for the most part, they are repeated sporadic communications in the lower left of the figure. These trajectories do not show the mobism of accumulating information as shown in Figs 8D and 8H, but are more extreme versions of the atomism shown in Figs 8B and 8F.

Thus, it was shown that dynamic changes in communication patterns can be typified based on the measures shown in Eqs (10)–(17) and in Table 3. It should be noted that none of the datasets searched for this time clearly indicated a situation of mobism, in which the community becomes enlarged or confused as information continues to increase. This may be because the purpose of collecting datasets was to analyze the network, and thus the datasets did not cover mobism, in which information continues to increase without changing the network structure. For this reason, we were unable to clarify the validity of the mobism measure $M_{mob} = \mu_L$ set in Table 3 in this study.

# Discussion

By capturing temporal changes in communication patterns on the network, as in this paper, rather than temporal changes in network structure as in conventional temporal network analysis, it is possible to identify communication features and typify the community. In real communities such as workplace, village, and conference, as opposed to special environments such as high school and primary school and virtual communities such as online community and email, the fact that the measure of mixism $M_{mix} = \mu_S \cdot \sigma_S^2$ balancing similarity and dissimilarity is large provides several suggestions.

Similarity is the "mixing" of humans with physical proximity, while dissimilarity is the "mingling" of diverse humans. These point to what philosopher Deguchi calls a "mixbiotic society" [7]. Being mixism also indicates that there is a balance between the generation and disappearance of communication, i.e., that the community is in a state of dynamic and living well-going.

The clear difference between real and virtual communities based on the presented measures anew reminds us of the importance of physical proximity and face-to-face contact. Communication through bidirectional, multimodal face-to-face contact is more important for mixbiotic and well-going communities than one-way, transient virtual communication.

The balance between passivity and active, synchrony and asynchrony according to CA's findings [42], and the balance between mixing (particle quantity) and blending (particle velocity) according to PRD's findings [45] are key to livingness (the edge of chaos). These findings suggest the importance of the balance between habituality and uncertainty, similarity and dissimilarity in communication in mixed societies. This is in line with the recognition that both the bonding and bridging types are important in the discussion of social capital (e.g. [75]). In the future, measuring the measure of mixism, $M_{mix} = \mu_S \cdot \sigma_S^2$, for various real communities will help to clarify the features and improvement policies of these communities.

With regard to virtual communities, we did not evaluate social networking services (SNS) with short messages and images in this study, but it is easy to imagine that the trend toward atomism will become even stronger in light of the results for online community and email. We can also predict to see the emergence of mobism-like phenomena such as those seen in SNS flare-ups. Although virtual communication is not expected to completely replace the role of face-to-face communication, its soundness will depend on the development such as the implementation of bidirectional and continuous communication that is not unidirectional or transient, the technology of multimodal communication, and the services combining real communication.

In this study, we used data on the time of sending and receiving messages in online community and email, but if there are replies to the same message, it can be considered that communication is continuing and information is accumulating. If we could collect data on replies in addition to sending and receiving times, we might be able to obtain different calculation results of measures and trajectories from this time. The same can be said for replies and thread continuation in SNS. Although online meeting and multiplayer video chat are considered relatively close to real communication, the issue is what range of communication time should be measured as substantive communication. This is because, unlike place-sharing in face-to-face contact, there are cases in which a meeting or chat screen is simply open. In addition, if the dynamic pattern of communication with avatars and AI in virtual space is similar to that of real communication, whether it is desirable as a mixbiotic society cannot be determined only by the measures set here, but will be necessary to be combined with other psychological measures and ethnomethodology, etc.

# Conclusions

As a contribution to sociology and network science, we proposed new mixbiotic society measures that indicates dynamic changes in communication patterns. Specifically, we first set the measures corresponding to the four phases of mixism (life, edge of chaos), atomism (poverty, isolation), mobism (hypertrophy or chaos), and nihilism (death) based on mathematical simulations of communication. Next, through validation with seven real-society datasets, we demonstrated the usefulness of the composite measure of mixism, $M_{mix} = \mu_S \cdot \sigma_S^2$, and the possibility of typifying communities based on the plural measures.

The mixbiotic society measures are superior to conventional social network analysis in that it can easily calculate temporal changes in communication on a network, it can be calculated without relying on a specific mathematical model compared to CA or PRD class decision, and the meaning of measures is easier to interpret than those. We believe that the mixism measure will be utilized as a measure of balanced well-going between similarity and dissimilarity, mixing and mingling, toward a desirable mixbiotic society.

As a future task, it is necessary to verify and review the validity of the atomism and mobism measures using datasets where atomism and mobism are actually occurring, as well as datasets related to SNS and social media. Further datasets would allow for correlation analysis with community features and principal component analysis for typologies. However, the perspectives for setting measures include simplicity of calculation, ease of interpretation of meaning, and ease of comparison among different communities. Introducing complex measures in order to increase precision and resolution, on the other hand, would compromise versatility, so we would like to consider this from multiple perspectives.

Note that the mixbiotic society measures do not analyze the network structure at each time, but rather analyze communication patterns on the network, and thus have the limitation of requiring network information over the all period beforehand, which at first glance do not seem suitable for real-time analysis. In practice, however, it is sufficient to calculate the measures from the start to the present time as needed, adding elements of the multidimensional vector of the information set as time passes from the start, and since the calculation load is not heavy, the analysis can be performed in quasi-real time. In addition, although time was counted as the number of times in the calculation of measures this time, a method of weighting the multidimensional vectors according to the duration of communication, for example, may be adopted.

Future developments will require validation with datasets followed by empirical research through fieldwork. The mixbiotic society measures can be described as objective measures of communication dynamics and livingness, but in combination with subjective measures such as, for example, the Self-as-WE scale [76] and the general well-being scale [77], which are appropriate for mixbiotic society. To lead societies and communities to well-going, we recommend that they be used comprehensively in the field. As candidates for the fieldwork, it would be interesting to target sites where social isolation and fragmentation are issues, as well as sites that span the real and virtual worlds of digital democracy [78] and platform cooperativism [79], where communication is critical.

# Acknowledgements

The authors received valuable comments on the perspectives of this study from the Hitachi Kyoto University Laboratory of the Kyoto University Open Innovation Institute. The authors would like to express their deepest gratitude. This work was supported by JST RISTEX Grant Number JPMJRS22J5, Japan.